\documentclass{aa}  
\usepackage{graphicx}
\usepackage{txfonts}
\usepackage{graphicx} \usepackage[final]{epsfig}
\usepackage{times}
\usepackage{latexsym}
\usepackage{amssymb}

\newcommand{\TMB}{\hbox {$T_{MB}$ }}
\newcommand{\cgn}{\hbox {CG~12-N }}
\newcommand{\cgnp}{\hbox {CG~12-N}}
\newcommand{\cgs}{\hbox {CG~12-S }}
\newcommand{\cgsp}{\hbox {CG~12-S}}

\newcommand{\twco}{{\hbox {\ensuremath{\mathrm{^{12}CO}} }}}
\newcommand{\twcop}{{\hbox {\ensuremath{\mathrm{^{12}CO}}}}}

\newcommand{\cso}{{\hbox {\ensuremath{\mathrm{C^{17}O}} }}}
\newcommand{\ceo}{{\hbox {\ensuremath{\mathrm{C^{18}O}} }}}

\newcommand{\thco}{{\hbox {\ensuremath{\mathrm{^{13}CO}} }}}

\newcommand{\htcopl}{{\hbox {\ensuremath{\mathrm{H^{13}CO^+}} }}}

\newcommand{\dcopl}{{\hbox {\ensuremath{\mathrm{DCO^+}}}\ }}
\newcommand{\dcoplp}{{\hbox {\ensuremath{\mathrm{DCO^+}}}}}

\newcommand{\kmps}{\ensuremath{\mathrm{km\,s^{-1}}}}
\newcommand{\Msun}{\ensuremath{\mathrm{M}_\odot}}
\newcommand{\kkms}{\ensuremath{\mathrm{K\,km\,s^{-1}}}}
\newcommand{\ms}{\ensuremath{\mathrm{m\,s^{-1}}}}
\newcommand{\hour}{\ensuremath{^\mathrm{h}}}
\newcommand{\minute}{\ensuremath{^\mathrm{m}}}

\newcommand{\Rthreetwo}{\ensuremath{\mathrm{R}_{32}}}

\newcommand{\Tex}{$T_{\rm ex}$}

\begin{document}

   \title{C$^{18}$O (3--2) observations of the Cometary Globule CG 12: A cold core and a \ceo hot spot}
%\thanks{Based 
 %on observations collected at the European Southern Observatory,  La Silla, Chile}}

   \author{L. K. Haikala\inst{1}
           \and M. Juvela\inst{1}
           \and J. Harju \inst{1}
           \and K. Lehtinen\inst{1}
           \and K. Mattila \inst{1}
           \and M. Dumke \inst{2}
          }

   \offprints{L. Haikala}

   \institute{Observatory, 
PO Box 14, University of Helsinki, Finland\\
              \email{haikala@astro.helsinki.fi}
         \and 
           European Southern Observatory, Casilla 19001, Santiago, Chile
}

   \date{Received ; accepted }

   \abstract{ The feasibility of observing the \ceo (3--2) spectral
line in cold clouds with the APEX telescope has been tested.  As the
line at 329.330 GHz lies in the wing of a strong atmospheric H$_2$O
absorption it can be observed only at high altitude observatories.
Using the three lowest rotational levels instead of only two helps to
narrow down the physical properties of dark clouds and globules.  The
centres of two \ceo maxima in the high latitude low mass star forming
region \object{CG 12} were mapped in \ceo (3--2) and the data were analyzed
together with spectral line data from the SEST.  The
\TMB(3--2)/\TMB(2--1) ratio in the northern \ceo maximum, \cgnp, is
0.8, and in the southern maximum, \cgsp, $\sim$2.  \cgn is modelled
as a 120\arcsec \ diameter (0.4pc) cold core with a mass of 27
\Msun. A small size maximum with a narrow, 0.8 \kmps, \ceo (3--2)
spectral line with a peak temperature of \TMB $\sim$ 11 K was detected
in \cgsp.  This maximum is modelled as a 60\arcsec$-$80\arcsec \
diameter ($\sim$0.2pc) hot (80 K $\lesssim$ \Tex $\lesssim$ 200 K)
$\sim$1.6 \Msun \ clump.  The source lies on the axis of a highly
collimated bipolar molecular outflow near its driving source. This is
the first detection of such a compact, warm object in a low mass star
forming region.
% If
%connected with the outflow,  the understanding  the formation and excitation
%mechanism of such objects may lead to a better understanding of the
%molecular outflow mechanism.

   \keywords{ clouds --  ISM  molecules -- ISM: structure -- 
              radio  lines -- ISM: individual objects: CG 12, NGC5367  }

   }

\titlerunning{C$^{18}$O (3--2) observations of the Cometary Globule CG 12}
   \maketitle
%
%________________________________________________________________

\section{Introduction}
%*********************************************************************
The two lowest rotational transitions of \twcop, \thco and \ceo have
been used extensively for studies of dark clouds and globules. Also
the $J=3-2$ lines of \twco and \thco can be routinely observed. In
contrast, there are very few published observations of the $J=3-2$
transition. Besides the nine point \ceo (3--2) map of S106 (Little et
al. 1995) and mapping of the centre of the pre-stellar cloud core
L1689B (\cite{jessop-WT}) only pointed observations are available
(e.g., J\o rgensen et al. 2002, 2004; Sch\"oier et al (2002); Lee et
al. 2003). The addition of the \ceo (3-2) transition to the lower 
transitions would, however,  substantially improve the accuracy of physical
parameters derived from \ceo data as the relative line intensities 
depend heavily on the cloud temperature  and density.

The SEST telescope has been  used to map numerous southern molecular clouds
in \ceo (1--0) and (2--1).  The new Atacama Pathfinder Experiment
(APEX\footnote{This publication is based on data acquired with the
Atacama Pathfinder EXperiment (APEX). APEX is a collaboration between
the Max--Planck--Institut f\"ur Radioastronomie, the European
Southern Observatory, and the Onsala Space Observatory.}) provides an
excellent opportunity to complement these data with \ceo (3-2)
observations.  The APEX beam size at 329 GHz, $\sim$19\arcsec, is
about equal to  the SEST beam size at the \ceo (2--1) frequency,
24\arcsec. This simplifies the comparison of APEX and SEST data.

In this Letter we demonstrate the use of \ceo (3-2), (2--1) and (1--0)
observations for radiative transfer modelling in the case of \object{Cometary
Globule 12}, CG 12, which is a high-latitude star forming region at a
distance of 630 pc (Williams et al. 1977). The head of the cloud
contains a highly collimated bipolar outflow (White et al. 1993) and
two compact \ceo maxima separated by about 3' (Haikala \& Olberg 2006,
Paper I). The centres of these maxima were mapped in \ceo (3--2).
Based on these observations a simple model is derived for both cores.

%__________________________________________________________________

\section{Observations and data reduction}   \label{sec:observations}

The observations were made with APEX  on Aug. 21
2005 in good weather (PWV 0.7 mm).  The \ceo (3--2) line at
329330.546 MHz was observed with the APEX-2A SIS DSB receiver. The
zenith optical depth was $\sim$ 0.27 and the DSB system temperature
ranged from 180 K to 280 K.  The receiver signal sideband lies in the
wing of an atmospheric H$_2$O absorption line whereas the mirror
sideband centered at 335.330 GHz is at a more transparent
frequency. The difference in the atmospheric opacities was estimated
using an atmospheric model and was taken into account in the
calibration.

Two 5 by 5 point maps with 10\arcsec \ spacing were obtained in the
position switching mode. Off positions (+5\arcmin \ in azimuth for
\cgn and +5\arcmin \ in azimuth and +5\arcmin \ in elevation for
\cgsp) were outside the globule at the time of the observations. \cgn
was observed at elevations $60\degr$ to $35\degr$ and \cgs at
$32\degr$ to $22\degr$. Integration time of 20s was used and
calibration was done every 10 to 30 minutes. Each position was
observed two or three times.  The map centre positions were integrated
for a total of 250s. The 1 GHz bandwidth of the MPIfR Fourier transform
spectrometer was divided into 16384 channels of 61 kHz ( $\sim$55\ms \
at 329 GHz).

 Most of the observed spectra contain ripple due to variations in the
atmospheric emission, reflections in the telescope optics and
instability of the receiver. In the data reduction the possible low
frequency ripple was first fit with a sinusoidal baseline where-after
possible higher frequency ripple was removed by masking the
corresponding frequency in Fourier transform space if possible. Finally a
first order baseline was fit to the spectra around the source velocity
and subtracted. The resulting rms of the spectra is by $\sim$15\%
higher than the value expected according to the radiometer formula
indicating that the possible non white noise due to, e.g., remaining
baseline ripple, does not dominate the spectral noise.  The spectra
were transformed to the main beam brightness temperature scale using a
main beam efficiency of 0.7. All the spectral line intensities and
line integrals reported in this paper are on the \TMB scale.

\section{Results}\label{sec:results}  

\begin{figure} \centering \includegraphics[width=8cm,angle=-90,clip]{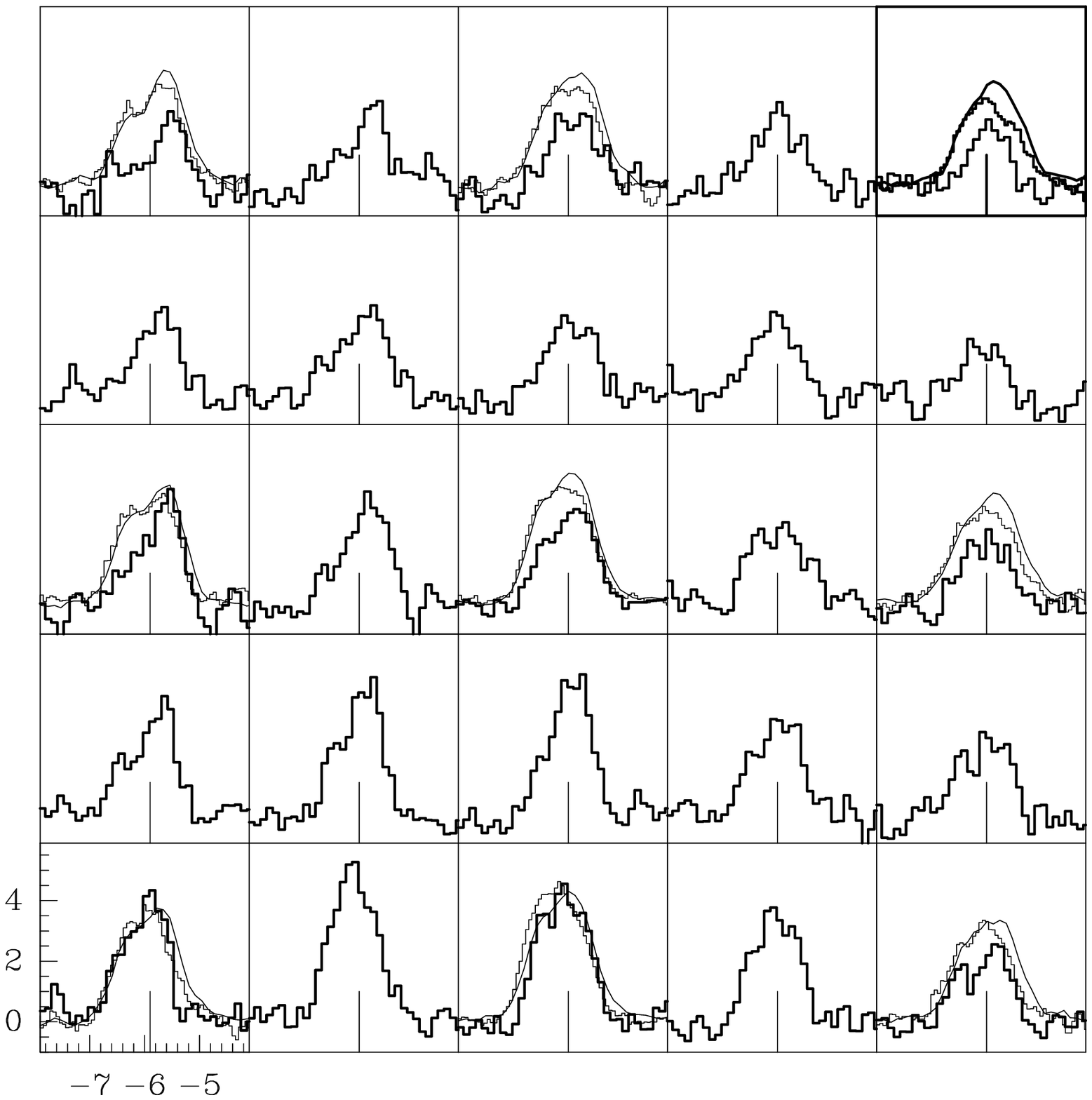}
 \caption{ The Apex \ceo (3--2) (hanning smoothed) and SEST \ceo (2--1)
and (1--0) (heavy histogram, light histogram and continuous line,
respectively) spectra  in \cgnp. The map grid step is
10\arcsec.  The tick mark is at velocity  $-$5.9 \kmps \ and the map centre
position is $13\hour 57\minute 39\fs5, -39\degr 56\arcmin 3\arcsec
$~(J2000).  }
\label{figure:c18on} 
\end{figure}

\begin{figure} \centering 
\includegraphics[width=8cm,angle=-90, clip]{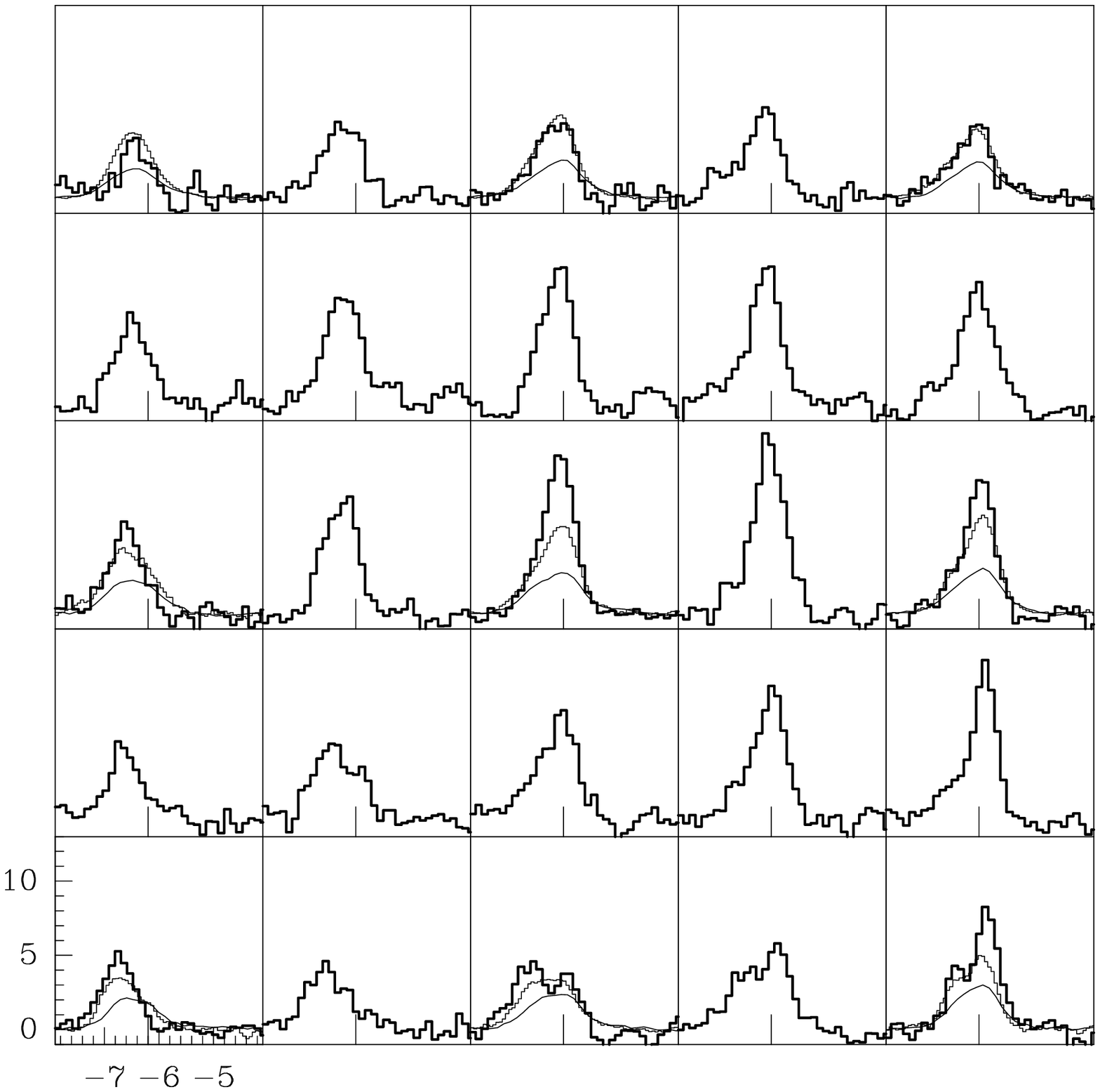}
\caption{ As Fig. \ref{figure:c18on} for \cgsp. The
tick line is at  $-$6.2 \kmps \ and the map centre
position is $13\hour 57\minute 41\fs4, -39\degr 59\arcmin 3\arcsec
$~(J2000). }
\label{figure:c18os} 
\end{figure}

\begin{figure*} \centering \includegraphics[width=7cm,angle=-90, clip]{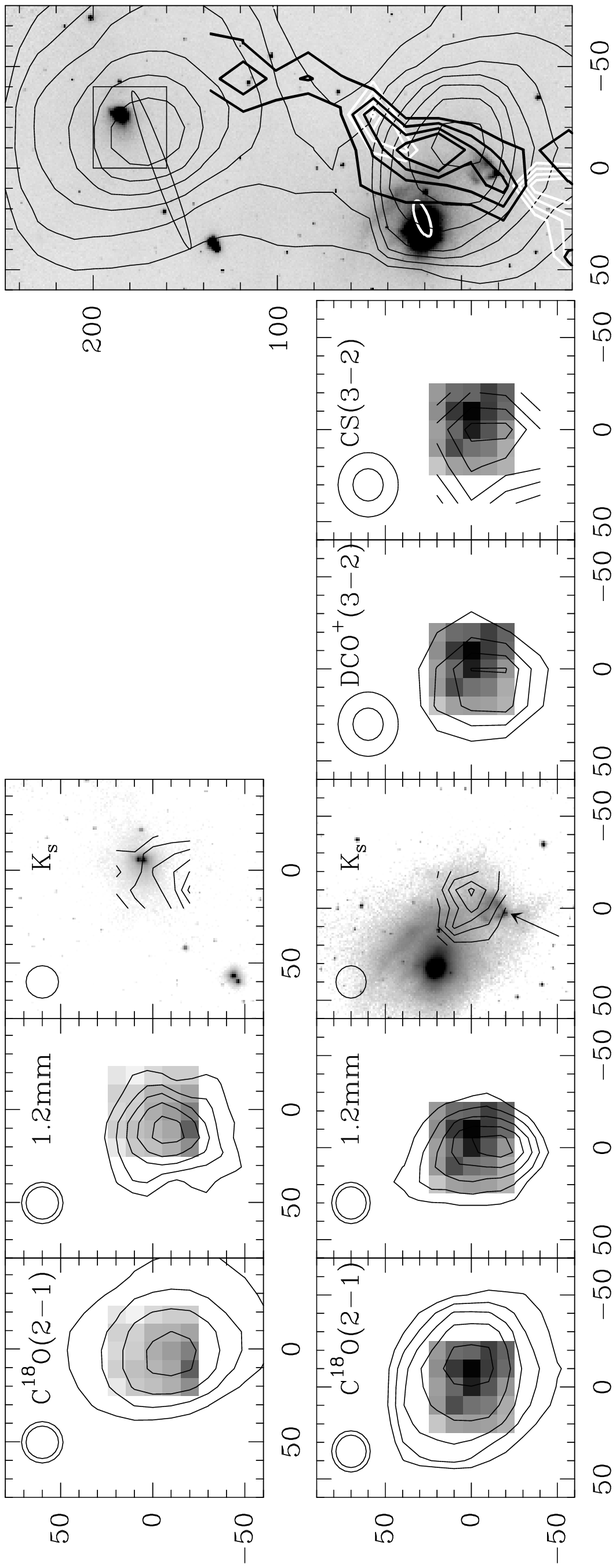}
\caption{ \ceo (3--2) line integral and SEST mm line and continuum
(arbitrary contour levels) and NTT NIR Ks observations. The upper row
shows \cgn and lower row \cgsp. \ceo (3--2) is shown in gray scale
except in the NIR Ks panels where contours are used. Lowest \ceo
(3--2) contour level and the increment are 2.0 \kkms \ and 0.6 \kkms \
for \cgn and 3.8 \kkms \ and 0.9 \kkms \ for \cgsp. The arrow in the
\cgs Ks panel shows the orientation of the outflow axis and points at
the assumed position of the driving source.  The APEX and SEST beam
sizes are shown in the upper left corner of each panel. The offset
from the map centres in Figs.  \ref{figure:c18on} and
\ref{figure:c18os} in arc seconds is shown on the axes. In the
rightmost panel the SEST \ceo (2--1) emission contours superposed on
the SOFI Ks image is shown. The lowest contour is 1.6 \kkms \ and the
increment is 0.6 \kkms. The
%origin is that of the \cgs \ceo (3--2) map and the box shows the
origin is that of Fig. \ref{figure:c18os} and the box shows the
location and size of the \cgn \ceo (3--2) map.  The red and blue outflow lobes from
\cite{white1993} are shown with heavy black and white contours,
respectively. The positional uncertainty ellipses of the IRAS point
sources \object{13547-3944} (in white) and \object{13546-3941} (in black) are also
shown.  }
\label{figure:molecules} 
\end{figure*}

\begin{figure} \centering \includegraphics[width=4.2cm,angle=-90, clip]{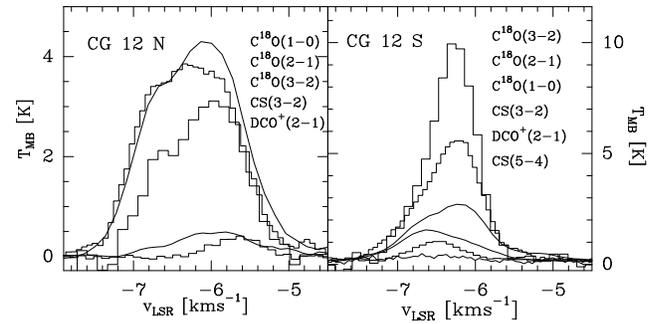}
\caption{\ceo, CS~(3--2) and \dcopl spectra in the \ceo (3--2) map
          centre positions. For \cgs also a CS(5-4) spectrum is
          shown }
\label{figure:spectra} 
\end{figure}

 The hanning smoothed \ceo (3--2) spectra in \cgn and \cgs are shown
in Figs.  \ref{figure:c18on} and \ref{figure:c18os} together with the
SEST \ceo (2--1) and (1--0) observations (Paper 1).  The area mapped
in \ceo (3--2) in each maximum covers only  the
SEST HPBW at 109 GHz, $\sim$47\arcsec, and the emission continues
beyond the map boundaries.
%Grey scale and contour 
Maps of the \ceo (3--2)  line integral 
are shown in Fig. \ref{figure:molecules} superposed on 
SEST  molecular line and  mm continuum 
maps  and a NIR Ks (NTT SOFI) image.

{\bf CG12-N:} In the southern part of the \cgn map the line profiles
and intensities of all the observed \ceo transitions are similar.
Elsewhere both the \ceo (3--2) line profile and line peak velocity
vary more than in the (2--1) transition.  The maximum is shifted to SE
from the \ceo (2--1) (and also (1--0)) maximum and is located near the
faint IRAS point source 12546-3941 (Fig. \ref{figure:molecules}).

The three \ceo transitions together with the CS (3--2) and \dcopl
(2-1) lines in the \cgn map centre position are shown in Fig.
\ref{figure:spectra}. The \ceo lines are asymmetric but the profiles
are similar.  An unpublished \cso (1--0) spectrum obtained in this
position shows that the \ceo (1--0) line is optically thin.  The weak
(\TMB$\approx 0.5$K) CS line peaks at the same velocity as the \ceo
lines. The \dcopl line is redshifted with respect to the \ceo and CS
lines.

{\bf CG12-S:} The \ceo (3--2) emission peaks 10\arcsec \ West of the
\cgs map centre position.   In contrast with \cgnp, the \ceo (3--2)
line is  significantly stronger than \ceo (2--1) in the map centre, 
West and South-West. 
The maximum lies  at the NW edge of the dense core outlined by the
high density tracers (\dcopl (2--1), CS (3--2)) and the 1.2 mm
continuum and on the axis of the outflow.
It is likely that
the driving source of the molecular outflow (White 1993) lies at the
Southern tip of the cone-like NIR nebulosity seen in the Ks
image at $+5\arcsec, -20\arcsec$. 
The nature of the driving source is not known.

In the \ceo (3--2) map centre position two \ceo velocity components
($-6.4$ \kmps and $-6.2$ \kmps, Paper 1) are blended and the lines are
skewed (Fig. \ref{figure:spectra}).  Only the $-6.4$ \kmps \ component
emits strongly in \dcoplp, CS (2--1), CS (3--2) and \htcopl (1--0)
(Paper 1).  However, in \ceo the $-6.2$ \kmps \ component is the
strongest. A two component gaussian fit with centre velocities fixed
to $-6.4$ \kmps \ and $-6.2$ \kmps \ gives 7.6 K, 3.7 K and 1.7 K,
respectively, for the \ceo (3--2), (2--1) and (1--0) peak temperatures
for the component centred at $-6.2$ \kmps. 
In \ceo (3--2) this component extends beyond
the mapped area in the South-West (Fig \ref{figure:c18os}).  The
source is elongated in the NE-SW direction but is hardly resolved
perpendicular to this direction.

\section{Discussion }\label{sec:discussion} 

The \TMB(3--2)/\TMB(2--1)  ratio (\Rthreetwo) is 
0.8 and 2 (the  $-6.2$ \kmps \ component) in \cgn and
\cgs, respectively. Assuming LTE, these ratios would correspond to
\Tex\  of $\leq$10 K in \cgn and $>$30 K in \cgs. 
Many of the \ceo line ratios in pre- and protostellar cores
(\cite{jorgensenetal2005} and references therein) are similar to those
observed in \cgnp. These line ratios were modelled with a ``drop''
model which assumes a strong molecular depletion in a cold shell ($<$30 K) 
between
the heating source  and the outermost part of the  envelope. 
%The dimensions of these sources are, however, smaller than that of \cgnp.
The intensity of the \ceo lines in IRAS 16293-2422 is as high as
observed in \cgs and \Rthreetwo \ is 1.3 (Sch\"oier et al, 2003).  The
lines are, however, optically thick and, e.g., the CS(5-4) line is
stronger than the \ceo (3--2) line. In \cgs the CS(5-4) line is not
detected (Fig. \ref{figure:spectra}).

 It is evident from Figs. \ref{figure:c18on} and \ref{figure:c18os}
that considerable fine structure in \ceo(3--2) is present both in
intensity and velocity in the two mapped areas. However, the signal-to-noise 
ratio of the spectra is not high enough for a more refined
analysis of the source structure than presented in Sect.
\ref{sec:results}. We therefore restrict our further analysis to
constructing  models which
reproduce the observed line ratios and intensities in the \ceo (3--2)
map central positions (Fig. \ref{figure:spectra}) where the signal-to-noise
ratio is best.  In \cgs only the $-6.2$ \kmps \ component
is considered.

 The reproducibility of the \ceo (1--0) and (2--1) spectral line
intensities at SEST has been very good. The lines were also observed
simultaneously and therefore the observed relative line intensities
should be quite accurate. No such record is available for the APEX
telescope and the receiver. We therefore assume a calibration accuracy
of 20\%.

Radiative transfer modelling was done with the Monte Carlo method
(Juvela 1997) assuming a spherically symmetric geometry. The C$^{18}$O
maps were used to estimate the cloud radii, 60\arcsec \ for \cgn and
35\arcsec \ for \cgsp. The  density
distribution adopted as a starting point was $n\propto
r^{-1.5}$ truncated at cloud radius. 
Power law temperature distributions were tested with
temperature either increasing or decreasing towards the cloud
centre. The model was fitted to the intensities and line widths of the
observed C$^{18}$O spectra (\cgn) and to line temperatures fitted
for the $-6.2$ \kmps \ component (\cgs).
The spectra around the central position were
used to constrain the radial density profile.  The free parameters of
the fit were the size of the inner region with constant density and
temperature, and the parameters of linear scalings that were applied
separately to the density and temperature values.  The model spectra
were convolved with gaussian beams that correspond to the resolution
of the observations. Constant C$^{18}$O abundance 
of 2.5 10$^{-7}$ was used. The quality of the fit depends only weakly on
the abundance which does, on the other hand, directly affect the mass
estimates.

The resulting radial density and temperature profiles are shown in
Fig. 5.  For \cgn the $\chi^2$-value is close to one. However, in the
case of \cgs the $\chi^2$-value is over five, mainly
because the predicted $J=1-0$ intensity is only half of  the
observed value. Best fits in \cgs were obtained with models where the kinetic
temperature increased inwards. One could obtain reasonable fits even
with isothermal models. \cgn is modelled with a 120\arcsec \ diameter
cloud where the temperature decreases from 12 K down to 6 K at the
edge.  The \cgs data can only be fit with a small size, hot
clump.  The modelled total gas column density in the APEX beam and the cloud
mass are $4.0 10^{22} {\rm cm}^{-2}$ and 27 \Msun \ for \cgn and $7.8
10^{21} {\rm cm}^{-2}$ and 1.6 \Msun \ for \cgsp, respectively.

 The models presented above should be considered only as
indicative.  Spherical 
%of the nature of these two \ceo maxima. 
symmetry was assumed and the \ceo (3--2) maps cover only the very
centres of the maxima.  Furthermore, the cloud sizes are close to the
telescope beam size so that the beam convolution has a large effect on
observed and calculated line ratios. Especially the \cgs fit is
sensitive to the beam convolution. Different combinations of source
size, temperature and density produce acceptable fits.  The central
density can not be increased much as the CS lines would become
stronger than observed. Common to all of these solutions is the small
size, a radius of  30\arcsec \ to 40\arcsec, and high central
temperature, 80 K to 200 K.  The problem is, however, not unique for
this source but inherent to all observing and modeling small size
(compared to beam size) sources.  The source beamfilling factor
becomes an important modelling parameter.

If the outer parts of the \cgn cloud envelope are heated up, e.g, by
interstellar radiation field (e.g., Snell 1981), this has very
small effect to the fit  at the cloud centre position. The gas
column  density in the cloud envelope is small compared to the
central density from where the bulk of the emission is coming.  The
LTE mass ( \Tex $=$ 10 K) calculated from the \ceo (2--1) data within
60\arcsec \ from the centre of \cgn is 25 \Msun \ which is in
agreement with the modeled mass of 27 \Msun.

The LTE mass of  the $-6.2$ \kmps \ velocity component in \cgs 
calculated from \ceo (2--1) data within 
35\arcsec \ from  the \ceo (3--2) map centre  is 4.1 \Msun \ for \Tex$=$15 K. 
However, for  \Tex$=$160 K the mass would be  three times higher.
The  LTE method assumes that all energy levels are thermalized  
and this is  clearly not the case in the \cgs model.

\begin{figure}\centering \includegraphics[width=4cm,angle=-90, clip]{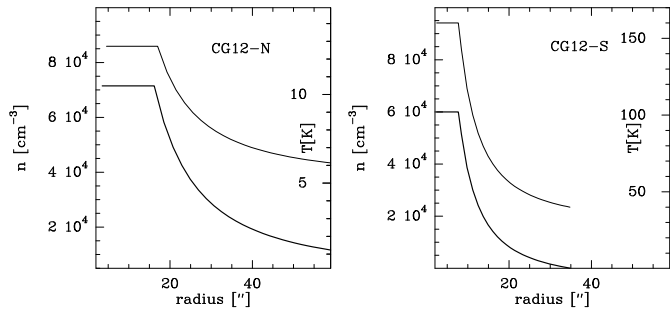}
 \caption{The radial density (lower curve) and temperature profiles of
\cgn and \cgsp. The horizontal axis shows the distance from the clump centre.
One arc-min corresponds to 0.18 pc or 3.8 $10^4$ AU at the 
assumed distance of 630~pc.} 
\label{figure:model} 
\end{figure}

\section{Conclusions}\label{sec:conclusions}

The addition of the \ceo (3--2) transition to the \ceo
data on the two lower transitions restricts strongly the parameter
space of possible cloud models.

The \ceo (3--2) observations of \cgn confirm the cold temperature of this
core. The \ceo (3--2) line profile varies more over the mapped
region than the \ceo (2--1) profile. The \ceo (3--2) line is therefore
probably better suited for studying the small-scale stucture of
molecular clouds than the lower transitions of this molecule.

 The detection of high, $\sim$10 K, \ceo (3--2) line temperatures in
\cgs was unexpected and forced us to reconsider the temperature and
density structure of the material traced by \ceo (2--1) and (1--0)
presented in paper 1.  The small, hot clump lies projected on the
edge of a dense core traced by \dcoplp. It also lies on the axis of a
highly collimated bipolar molecular outflow near its driving
source. This suggests that the two are related. As the \ceo lines from
the hot spot are narrow and appear at the same radial velocity as the
parent cloud, they probably do not originate in shocked gas. Neither
is direct radiative heating from the strong IRAS point source
13547-3944 likely. The heating mechanism and the relation of the hot
spot to the collimated outflow remain therefore unspecified.  Further
studies of this phenomenom seem warranted as they may lead to a better
understanding of the interaction between newly born stars and the
surrounding molecular material.

Better spatial resolution can be achieved  by
observing still higher \ceo transitions
which should be detectable
in \cgsp. The sizes of the CG 12 \ceo maxima are well suited for the
future ALMA interferometer

%\listofobjects

\end{document}